\documentclass{iopart}
\usepackage{iopams,graphicx}

\begin{document}

\letter{Spin exchange in quantum rings and wires in the
Wigner-crystal limit}

\author{Michael M Fogler and Eugene Pivovarov}

\address{Department of Physics, University of
California San Diego, La Jolla, California 92093, USA}
\ead{mfogler@ucsd.edu}

\begin{abstract}
We present a controlled method for computing the exchange coupling in
strongly correlated one-dimensional electron systems. It is based on the
asymptotically exact relation between the exchange constant and the
pair-correlation function of spinless electrons. Explicit results are
obtained for thin quantum rings with realistic Coulomb interactions, by
calculating this function via a many-body instanton approach.
\end{abstract}


\submitto{\JPCM}


\section{Introduction}

Much attention has been devoted to the spin degree of freedom in
one-dimensional (1D) conductors, both of the linear shape (quantum
wires~\cite{Auslaender_05}, carbon nanotubes~\cite{Jarillo-Herrero_04})
and of a circular one (quantum rings~\cite{Lorke_00, Fuhrer_04,
Bayer_03}). Physical parameters of such systems, e.g., average distance
between the electrons $a$, their total number $N$, their effective mass
$m$, dielectric constant $\epsilon$, etc., can vary over a broad
range or can be tuned experimentally. This creates unique opportunities
for studying the effect of reduced dimensionality and strong Coulomb
interactions on quantum magnetism. A number of new developments have
generated a particular interest to the physics of a 1D Wigner-crystal
(WC). Unlike the case in higher dimensions, in 1D the crossover to this
strongly-correlated regime occurs at easily achievable electron
densities~\cite{Egger_99}, $r_s \equiv a / 2 a_B > 4$, where $a_B =
\hbar^2 \epsilon / m e^2$ is the effective Bohr radius. Disorder has
been the only major obstacle to realizing the 1D WC
experimentally~\cite{Auslaender_05}. A
successfully solution to this problem has been apparently found, at
least, for the case of carbon nanotubes. Very large $r_s$ values have
been recently demonstrated in suspended nanotube devices without
appreciable intervention of disorder effects~\cite{Jarillo-Herrero_04}.
Because of their finite length, in the desired range $r_s > 4$ these
devices contained only a few electrons, $N < 25$. Such finite-size
systems are traditionally referred to as Wigner
molecules~\cite{Reimann_02}. The progress towards realizing Wigner
crystal (molecule) states in GaAs quantum wires has also been very
encouraging~\cite{Auslaender_05}, therefore, one may hope that they will
soon follow suit.

\begin{figure}
\begin{center}
\includegraphics[height=1.75in]{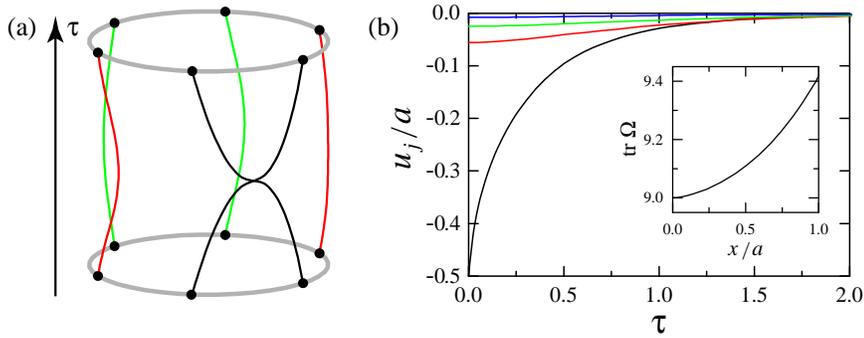}
\end{center}
\caption{(Colour online)
The instanton trajectories. (a) Schematic representation for $N = 6$
Wigner molecule on a ring. (b) The trajectories of $1\leq j\leq 4$
electrons for $N = 8$ (for notations used see the main text). Inset:
function $\tr\bOmega(x)$. The units of $\tau$ and $\bOmega$ are
$\sqrt{2} a / s_0$ and its inverse.
\label{fig:Instanton}}
\end{figure}

On the theoretical side, the 1D WC is interesting because of a dramatic
difference between the characteristic energy scales for orbital and spin
dynamics. This \emph{strong\/} spin-charge separation has been recently
predicted to cause anomalies in many essential electron properties,
e.g., ballistic conductance~\cite{Matveev_04} of quantum wires and
persistent current of quantum rings~\cite{Reimann_02}. In view of the
above, obtaining a reliable estimate of the spin-related energy scales,
notably, the exchange-coupling $J$ of the nearest-neighbour electrons, is
desirable. It has been an outstanding challenge, though. As depicted in
\fref{fig:Instanton}(a), $J$ is determined by the acts of quantum
tunnelling in which any two such electrons interchange. At $r_s \gg 1$
the corresponding potential barrier greatly exceeds the kinetic energy
of the electron pair, which makes $J$ exponential small and difficult to
compute numerically~\cite{Reimann_02}. Attempts to derive $J$
analytically (for the nontrivial case $N > 2$) were based on the
approximation that neglects all degrees of freedom in the problem except
the distance between the two interchanging electrons~\cite{Matveev_04,
Hausler_96}. We call this a Frozen Lattice Approximation (FLA). The
accuracy of the FLA is unclear because it is not justified by any small
parameter. When a given pair does its exchange, it sets all other
electrons in motion, too (\fref{fig:Instanton}). To obtain the much
needed reliable estimate of $J$ one has to treat the spin exchange in a
Wigner molecule (or a WC) as a truly many-body process. This is done
below in this letter, where we compute $J$ to the leading order in the
large parameter $r_s$.

%


\section{Model and results}

We assume that electrons are tightly confined in the transverse
dimensions on a characteristic lengthscale $R \ll a_B$. This allows us
to treat the problem as strictly 1D, provided we replace the Coulomb law
by the appropriate effective interaction that goes to a finite value at
distances $r \ll R$. We adopt a simple model form~\cite{Fogler_05,
Usukura_05} $U(r) = e^2 / \epsilon (r + R)$, which is the simplest
expression that correctly captures both short- and long-range behaviour
of the (unscreened) Coulomb interaction for \emph{any\/} realistic
confining potential and is similar to other forms used in the
literature~\cite{Friesen_80, Szafran_04}. For convenience, we focus on
the quantum ring geometry where $r = (N a / \pi) \left|\sin (\pi x / N
a)\right|$ is the chord distance and $x$ is the coordinate along the
circumference.

In the Wigner molecule configuration electrons reside
at the corners of a regular polygon. The effective low-energy
Hamiltonian of such a state is given by~\cite{Reimann_02}
\begin{equation}
H = \frac{\hbar^2}{2 I} L^2
  + J \sum_{j} \mathbf{S}_j \mathbf{S}_{j + 1}
  + \sum_{\alpha} n_a \hbar \omega_{\alpha},
\label{eqn:H}
\end{equation}
where $L$ is the centre-of-mass angular momentum, $\mathbf{S}_j$ are
electron spins and $n_{\alpha}$ are the occupation numbers of
``molecular vibrations.'' At large $r_s$, the total moment of inertia
$I$ and the vibrational frequencies $\omega_{\alpha}$ are easy to
compute because they are close to their classical values. In order to
calculate $J$, which is more difficult, we first show that the
asymptotically exact relation exists between $J$ and the
pair-correlation function (PCF) $g(x)$ of a \emph{spin polarized\/} 1D
system. For an ultrathin wire, $\mathcal{L} \equiv \ln ({a_B}/{R}) \gg
1$, it is particularly compact:
\begin{equation}
 J = \left( {e^2 a_B^2} / {2 \mathcal{L} \epsilon}\right)
    g^{\prime\prime}(0),\quad r_s \gg 1 / \mathcal{L}.
\label{eqn:J_from_g_ultrathin}
\end{equation}
By virtue of \eref{eqn:J_from_g_ultrathin}, the calculation of $J$
reduces to an easier task of computing $g(x)$. Using the instanton
method described below we arrive at the final result
\begin{equation}
J = \frac{\kappa}{\left( 2 r_s\right) ^{5/4}} \frac{\pi}{\mathcal{L}}
    \frac{e^2}{\epsilon a_B}
    \exp \left( -\eta \sqrt{2 r_s}\,\right),
    \quad r_s \gg 1.
\label{eqn:J}
\end{equation}
The values of $\eta$ and $\kappa$ are given in \tref{tbl:Results}. They
demonstrate that the FLA~\cite{Matveev_04, Hausler_96} errs
significantly in $\kappa$, by about $50\%$, but surprisingly little in
$\eta$, only by 0.7\%.

\begin{table}
\caption{\label{tbl:Results}Results for Wigner molecules on a ring
(finite $N$) and for wires ($N=\infty$).}
\begin{indented}
\item[]\begin{tabular}{lcccccc}
\br
$N$      &3      &4         &6         &8         &$\infty$  &$\infty$-FLA\\
\mr
$\eta$   &2.8009 &2.7988(2) &2.7979(2) &2.7978(2) &2.7978(2) &2.8168      \\
$\kappa$ &3.0448 &3.18(6)   &3.26(6)   &3.32(7)   &3.36(7)   &2.20        \\
\br
\end{tabular}
\end{indented}
\end{table}
%


\section{Three electrons on a ring}

We start with the simplest nontrivial case: $N = 3$. Let $0 \leq x_j
< 3 a$, $j = 0, 1, 2$, be the electron angular coordinates. We will
compute the exchange coupling $J$ between the $j = 0$ and the $j = 1$
electrons. It is convenient to go to new variables: the relative
distance of the pair, $x \equiv x_1 - x_0$ and the location
of the $j = 2$ electron with respect to the centre of mass $X_{2} \equiv
x_2 - x_{\mbox{cm}} - a$. The motion of the centre of mass
$x_{\mbox{cm}}$ can be ignored. We arrive at the problem with one
fast ($x$) and one slow ($X_{2}$) degree of freedom. (Classically,
$X_{2} = 0$.) The total potential energy $U_{\mbox{tot}}(x, X_2) = U(x)
+ U\left[(3 / 2) (X_2 + a) - x / 2\right] + U\left[(3 / 2) (X_2 + a) + x
/ 2\right]$ has two global minima in the fundamental domain $|x| < 3/2
a$, at $x = \pm a$, $X_{2} = 0$. They give rise to the two lowest-energy
multiplets: the spin-singlet ground state $\mathbf{S}_0 + \mathbf{S}_1 = 0$
with an orbital wavefunction $\Phi_s(x, X_2)$ and the triplet with
a wavefunction $\Phi_t$. Their energy splitting is the desired exchange
coupling $J$. It is given by the formula~\cite{Herring_64,
Landau_III}
\begin{equation}
 J = (2 \hbar^2 / \mu) \int d X_2
     \left.\Phi_1 \partial_x \Phi_1\right|_{x = 0},
\label{eqn:J_from_Phi_1}
\end{equation}
where the (normalized to unity) ``single-well'' wavefunction $\Phi_1(x,
X_2)$ is the ground-state of the Hamiltonian with a modified potential
$U_{\mbox{tot}} \to U_1 \equiv U_{\mbox{tot}}\left(\max\{x, 0\}, X_2\right)$
and $\mu = m/2$. \Eref{eqn:J_from_Phi_1} is valid to order
$O(J^2)$~\cite{Herring_64}; with the same accuracy, the singlet and triplet
wavefunctions are symmetric and antisymmetric combinations of the
single-well wavefunctions,
$\Phi_{s,t} = \left[\Phi_1(x, X_{2}) \pm \Phi_1(-x, X_{2})\right] / \sqrt{2}$.

Let us discuss the form of $\Phi_1(x, X_{2})$. Near its maximum at
$x = a$, $X_2 = 0$, it is a simple Gaussian in both variables,
characterized by an amplitude $l$ of the zero-point motion in $x$
and a frequency $\Omega(a)$ of the zero-point oscillations in $X_{2}$.
Away from its maximum $\Phi_1$ rapidly decays at
$|X_2| \gtrsim l \gg a$. This justifies the following Gaussian
approximation\footnote{The
Gaussian ansatz has been used previously for computing
spin exchange in $^3${\sc H}e crystals~\cite{Roger_83}.} in the
\emph{entire\/} fundamental domain of $x$:
\begin{equation}
\Phi_1 = \phi(x) \exp\left[-(M / 2 \hbar) \Omega(x) X_2^2\right] ,
\label{eqn:Phi_1_3}
\end{equation}
where $M = 3\mu$.
It is important that at $x \ll a$, where the tunnelling barrier is
large, $\Omega$ is a slow function of $x$. Hence, if $g(x)$ denotes the
PCF of a spin-polarized molecule,
\begin{equation}
  g(x) \equiv 2 \int \prod _{j = 2}^{N - 1} d X_j
              \Phi_t^2\left(x, X_2,\ldots, X_{N - 1}\right),
\quad |x| < {3 a / 2},
\label{eqn:g_def}
\end{equation}
then \eref{eqn:J_from_Phi_1} immediately entails
$J = \left({\hbar^2} / {4 \mu}\right)
     \left[{\phi(0)} / {\phi^\prime(0)}\right]
     g^{\prime\prime}(0)$.
Anticipating the discussion below, \eref{eqn:g_def} is written for
an arbitrary $N > 2$, with the notation $X_j \equiv x_j - x_{\mbox{cm}} +
(N - 1 - 2 j) (a / 2)$ being used; the PCF is normalized as
appropriate in the WC limit, $\int_{0}^{3a/2} g(x)\, dx = 1$.

%
\begin{figure}
\begin{center}
\includegraphics[height=1.3in]{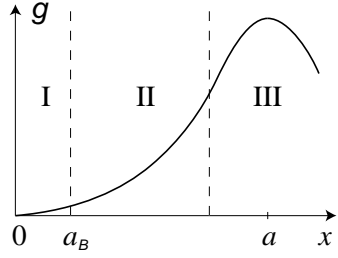}
\end{center}
\caption{The PCF of a spin-polarized system (schematically). Regions I, II
and III are described in the main text.
\label{fig:PCF}}
\end{figure}

The equations on $\phi(x)$ and $\Omega(x)$ are obtained by
substituting \eref{eqn:Phi_1_3} into the Schr\"odinger equation and
neglecting terms small in $l/a$. This results in
the dependence of $g(x)$ on $x$ sketched in
\fref{fig:PCF}. Near its $x = a$ maximum (region III) $g(x)$ is
a Gaussian of width $l$. In the region II the quasiclassical approximation applies.
Finally, in an ultrathin wire, $\mathcal{L} \gg 1$, there is also
region I, $x \lesssim a_B$, where the
quasiclassical approximation breaks down. Fortunately,
the equations on $\phi(x)$ and $\Omega(x)$ can be simplified there, as
$\Omega(x)\simeq\Omega(0)$ and $U_{\mbox{tot}}(x)\simeq U(x)+2U(3a/2)$.
Similar to~\cite{Fogler_05}, this leads to
$\phi(0)/\phi^\prime(0)\simeq a_B /\mathcal{L}$,
which, combined with the expression for $J$,
yields equations (\ref{eqn:J_from_g_ultrathin})
and (\ref{eqn:J}), with $\eta$ and $\kappa$ given by
\begin{eqnarray}
\eta &=& 2 \int\limits_0^a \frac{d x}{a}
      \left[\frac{\epsilon a}{e^2}\, \Delta U_{\mbox{tot}}(x)
      \right]^{1/2},
\label{eqn:eta_3}\\
\kappa &=& \frac{2^{5 / 4}}{\sqrt{\pi}}\, e^{\xi(0)}
          \sqrt{\frac{\Omega(a)}{\Omega(0)}}
          \left[
            \frac{\epsilon a^3}{e^2}\, U^{\prime\prime}_{\mbox{tot}}(a)
          \right]^{3/4}.
\label{eqn:kappa_3}
\end{eqnarray}

Thus, for the $N = 3$ case we were able to reduce the original
complicated three-body eigenvalue problem to routine operations of
solving an ordinary differential equation on $\Omega(x)$
and taking two quadratures.
The resultant $\eta$ and $\kappa$ are listed in \tref{tbl:Results}.
In comparison, the FLA~\cite{Matveev_04} underestimates $\kappa$ by about
$50\%$. It gets $\eta$ correctly but only for $N = 3$, see more below.

One important comment is in order. The antisymmetry of the total fermion
wavefunction imposes certain selection rules~\cite{Maksym_96} for the
allowed values of $L$ [see \eref{eqn:H}] at a given total spin $S$.
The lowest-energy $L$ eigenstates for the two possible $S$ values in the
$N = 3$ system, $S = 1/2$ and $S = 3/2$, are $|L| = 1$ and $0$,
respectively. Since $J \ll \hbar ^{2}/I$ at large $r_s$, the ground
state of the system is the $L = 0$ spin-quartet~\cite{Reimann_02,
Usukura_05}.


\section{$N > 3$ electrons on a ring}

In a system of more than three electrons, the single-well function
$\Phi_1(x, \mathbf{X})$ can be sought in the form similar to
\eref{eqn:Phi_1_3}, but with the argument of the exponential replaced by
$(-1/2\hbar) (\Delta\mathbf{X}^\dagger \mathbf{M}^{1/2}) \bOmega(x)
(\mathbf{M}^{1/2}\Delta\mathbf{X})$, where $M^{-1/2}_{ij} =
m^{-1/2}[\delta_{ij}- (1 - \sqrt{2 / N}) / (N - 2)]$. In the language of
the quantum tunnelling theory, $\bOmega(x)$ is a matrix that controls
Gaussian fluctuations $\Delta \mathbf{X} = \mathbf{X} - \mathbf{X}^\ast$
around the instanton trajectory $\mathbf{X}^\ast(x)$, where $\mathbf{X}
= (X_2,\ldots, X_{N - 1})^{T}$. Switching to the usual
parametrization of the instanton by an ``imaginary-time'' $\tau$, we
seek $x(\tau)$ and $\mathbf{X}^\ast(\tau)$ that minimize the action
\begin{equation}
S_N = \int\limits_0^\infty \frac{d \tau}{\hbar}
    \left[\frac{\mu}{2} (\partial_\tau x)^2 +
          \frac12 (\partial_\tau \mathbf{X})^\dagger \mathbf{M}\,
                   \partial_\tau \mathbf{X}
    + \Delta U_{\mbox{tot}}
    \right] ,
\label{eqn:S_N}
\end{equation}
subject to the boundary conditions $x(0) = 0$, $x(\infty) = a$ and
$\mathbf{X}(\infty) = 0$. Henceforth $U_{\mbox{tot}}$ is always meant to
be evaluated on the instanton trajectory and $\Delta U_{\mbox{tot}}$
stands for the difference of its values at a given $\tau$ and at $\tau =
\infty$. Repeating the steps of the derivation for the $N = 3$ case, we
derive the following equations on $\phi(x)$ and $\bOmega(x)$:
\begin{equation}
\eqalign{
\partial_\tau \bOmega = \bOmega^2(\tau) - \bomega^2(\tau), \cr
\left\{(\hbar^{2} / 2\mu) \partial^2_x - U_{\mbox{tot}}(x)
 - (\hbar / 2) \tr\,\bOmega(x) + E\right\} \phi(x) = 0,}
\label{eqn:Omega}
\end{equation}
where $\bomega$ is a positive-definite matrix such that
$\bomega ^2=\mathbf{M}^{-1/2}\bXi \mathbf{M}^{-1/2}$ and
$\bXi$ is the matrix of the second derivatives $\Xi_{i j} =
\partial_{X_i} \partial_{X_j} U_{\mbox{tot}}$.
The equations are mutually
consistent if $E = U_{\mbox{tot}}(a) + (\hbar / 2) [\tr\bomega(a) +
\omega_0]$, $\omega_0 \equiv \hbar / \mu l^2$ and $\bOmega(a) =
\bomega(a)$.

The PCF $g(x)$ in the quasiclassical region can be written in terms of
the tunnelling action (\ref{eqn:S_N}) and the appropriate prefactor
as follows:
\begin{eqnarray}
g(x) &=& \frac{a}{l^2}
     \left[\frac{1}{2\pi} \frac{\Omega(a)}{\Omega(x)}
           \frac{\hbar\omega_0}{U(x)}
     \right]^{1/2}
     e^{\xi(x) - 2 S_N(x)},
\label{eqn:g_II}\\
\xi(x) &=& \int\limits_x^a d y \left\{
          \frac{\omega_0 + \tr\bOmega(a) - \tr\bOmega(y)}
               {\left[(2 / \mu) \Delta U_{\mbox{tot}}(y)\right]^{1/2}}
            - \frac{1}{a - y}
           \right\}.
\label{eqn:xi}
\end{eqnarray}
Here the action $S_N$ is defined to be the value of the integral in
\eref{eqn:S_N} when its lower limit is replaced by $\tau = \tau(x)$. For
$\eta$ we find $\eta = 2 S_N / \sqrt{2 r_s}$, while $\kappa$
is given by equation (\ref{eqn:kappa_3}) after the replacement $\Omega
\to \det\bOmega$.


\section{Calculation of the instanton}

A few properties of the instanton follow from general considerations.
The dimensional analysis of action (\ref{eqn:S_N}) yields $S
_N\propto\sqrt{r_s}$, so that $\eta$ is indeed just a constant. Also,
from the symmetry of the problem, $X_{N + 1 - j}(\tau) = -X_j(\tau)$.
Thus, in the special case of $N = 3$, the instanton trajectory is
trivial: $X_2 \equiv 0$, i.e., the $j = 2$ electron does not move. This
is why we were able to compute $S_3$ in a closed form. For $N > 3$ the
situation is quite different: all electrons [except $j = (N + 1) / 2$
for odd $N$] \emph{do\/} move. In order to investigate how important the
motion of electrons distant from the $j = 0, 1$-pair is let us consider
the $N = \infty$ (\emph{quantum wire\/}) case, where the far-field
effects are the largest. If $X_j$'s were small, we could expand $\Delta
U_{\mbox{tot}}$ in \eref{eqn:S_N} to the second order in $X_j$ to obtain
the harmonic action
\begin{equation}
S_h = \frac12 \frac{m}{\hbar} \int
     \frac{d k}{2\pi}
     \int \frac{d\omega}{2\pi} \left|u_{k \omega}\right|^2
     \left[\omega^2 + \omega_p^2(k)\right],
\label{eqn:action_harmonic}
\end{equation}
where $u_{k \omega}$ is the Fourier transform of electron displacement
$u_j(\tau) \equiv x_j - x_j^0$ from the classical equilibrium position
$x_j^0 \equiv (j - 1 / 2) a$, $j \in \mathbb{Z}$, $\omega_p(k) \simeq
s_0 k \ln^{1/2}(4.15 / k a)$ is the plasmon dispersion in the 1D WC and
$s_0 \equiv (e^2 / \epsilon\mu a)^{1/2}$. Minimization of $S_h$ with the
specified boundary conditions yields $u_j(\tau) \propto v x_j^0
/\left[(x_j^0)^2 + v^2 \tau^2\right]$, where $v \simeq (s_0 / 2)
\ln\left\{\left[(x_j^0)^2 + s_0^2 \tau^2\right] / a^2\right\}$.
Substituting this formula into harmonic action
(\ref{eqn:action_harmonic}), we find that the contributions of distant
electrons to $S_h$ rapidly decay with $|j|$. Thus, a fast convergence of
$\eta$ to its thermodynamic limit is expected as $N$ increases.
Encouraged by this conclusion, we undertook a direct numerical
minimization of $S$ for the set of $N$ listed in \tref{tbl:Results}
using standard algorithms of a popular software package \textsc{MATLAB}.
The optimal trajectories that we found for the case of $N = 8$ are shown
in \fref{fig:Instanton}(b). As one can see, electron displacements
reach some finite fractions of $a$ at $\tau = 0$. This collective
electron motion lowers the effective tunnelling barrier and causes $\eta$
to drop below its FLA value, although only by $0.7\%$, see
\tref{tbl:Results}.

Let us now discuss the prefactor $\kappa$. In the inset of
\fref{fig:Instanton}(b) we plot $\tr\bOmega(x)$
computed by solving \eref{eqn:Omega} numerically. To reduce the
calculational burden, we set $\mathbf{X}^\ast(\tau) \to 0$ instead of
using the true instanton trajectory. The error in $\kappa$ incurred
thereby is $\sim 2\%$. In comparison, the FLA, where
$\tr\bOmega(x) = \mbox{const}$, yields $\kappa$ about
$50\%$ smaller than the correct result, similar to $N = 3$.


\section{Relation to current experiments}

For carbon nanotube quantum dots~\cite{Jarillo-Herrero_04}, where the WC
limit has apparently been realized, our formula~\eref{eqn:J} gives $J
\sim 1\,{\rm K}$ at $r_s = 4$, which should be verifiable
experimentally. Unfortunately, the lowest measurement temperature was
$0.3\,{\rm K}$; therefore, the exchange correlations may have been
washed out. We hope that our predictions can be checked in the next
round of experiments. Energy-level spectroscopy of quantum
rings~\cite{Lorke_00, Fuhrer_04, Bayer_03} is another promising area
where our results may apply. In longer 1D wires, $J$ determines the
velocity $v_\sigma = (\pi / 2) J a / \hbar$ of spin excitations, which
can be measured by tunnelling~\cite{Auslaender_05},
photoemission~\cite{Claessen_02}, or deduced from the enhancement of the
spin susceptibility and electron specific heat~\cite{Fogler_05}. Our
result for $v_\sigma$ reads (cf. \tref{tbl:Results})
\begin{equation}
v_{\sigma}/v_{F} = 5.67 \left( \pi / \mathcal{L}\right)
r_s^{3/4} e^{-\eta\sqrt{2 r_s}},\quad \eta = 2.7978(2),
\label{eqn:v_sigma_Coulomb}
\end{equation}
where $v_F = (\pi / 2) (\hbar / m a)$ is the Fermi velocity.


\bigskip\noindent\ignorespaces
This work is supported by the A.~P. Sloan and the C.~\&~W.
Hellman Foundations.


\bigskip\noindent\emph{Note added.}---After the completion of this work, we
learned that Klironomos \etal \cite{Klironomos_05} independently computed
$\eta = 2.79805(5)$, but not the prefactor $\kappa$. These authors also
considered a correction to $\eta$ due to a finite radius of the wire
$R$. We can show that as $R$ increases, the ratio $\pi / \mathcal{L}$ in
\eref{eqn:v_sigma_Coulomb} is replaced by a more complicated
expression that tends to unity at $R > a_B$.



\section*{References}
\begin{thebibliography}{99}

\bibitem{Auslaender_05} Auslaender O~M, 
 Steinberg~H, Yacoby~A, Tserkovnyak~Y, Halperin B~I,
 Baldwin K~W, Pfeiffer L~N and West K~W
2005 \textit{Science} \textbf{308} 88
and references therein
\item[] Field S~B, Kastner M~A, Meirav~U,
 Scott-Thomas J~H~F, Antoniadis D~A, Smith H~I and Wind S~J
1990 \PR B \textbf{42} 3523

\bibitem{Jarillo-Herrero_04} Jarillo-Herrero~P, 
 Sapmaz~S, Dekker~C, Kouwenhoven L~P and van der Zant H~S~J
2004 \textit{Nature} \textbf{429} 389

\bibitem{Lorke_00} Lorke~A, 
 Luyken R~J, Govorov A~O, Kotthaus J~P,
 Garcia J~M and Petroff P~M
2000 \PRL \textbf{84} 2223
\item[] Warburton R~J, 
 Sch\"aflein~C, Haft~D, Bickel~F, Lorke~A,
 Karrai~K, Garcia J~M, Schoenfeld~W and Petroff P~M
2000 \textit{Nature} \textbf{405} 926

\bibitem{Fuhrer_04} Fuhrer~A, Ihn~T, Ensslin~K, Wegscheider~W
and Bichler M
2004 \PRL \textbf{93} 176803

\bibitem{Bayer_03} Bayer~M, 
 Korkusinski~M, Hawrylak~P, Gutbrod~T, Michel~M
 and Forchel~A
2003 \PRL \textbf{90} 186801

\bibitem{Egger_99} Egger R, H\"ausler W, Mak C~H and Grabert~H
1999 \PRL \textbf{82} 3320

\bibitem{Reimann_02} Reimann S~M and Manninen~M
2002 \RMP \textbf{74} 1283
\item[] Viefers~S, Koskinen~P, Singha Deo~P and Manninen~M
2004 \textit{Physica} E \textbf{21} 1

\bibitem{Matveev_04} Matveev K~A
2004 \PR B \textbf{70} 245319

\bibitem{Hausler_96} H\"ausler~W
1996 \ZP B \textbf{99} 551

\bibitem{Fogler_05} Fogler M~M
2005 \PR B \textbf{71} 161304(R)

\bibitem{Usukura_05} Usukura~J, Saiga~Y and Hirashima D~S
2005 \JPSJ \textbf{74} 1231
and references therein

\bibitem{Friesen_80} Friesen W~I and Bergerson B
1980 \JPC \textbf{13} 6627

\bibitem{Szafran_04} Szafran~B, Peeters F~M,
Bednarek~S, Chwiej~T and Adamowski~J
2004 \PR B \textbf{70} 035401

\bibitem{Herring_64} Herring~C
1964 \PR \textbf{134} A362

\bibitem{Landau_III} Landau L~D and Lifshitz E~M
1977 \textit{Quantum Mechanics} (Oxford: Pergamon) sec 81

\bibitem{Roger_83} Roger~M, Hetherington J~H and Delrieu J~M
1983 \RMP \textbf{55} 1

\bibitem{Maksym_96} Maksym P~A
1996 \PR B \textbf{53} 10871
\item[] Koskinen~P, Koskinen~M and Manninen~M
2002 \textit{Eur. Phys. J.} B \textbf{28} 483

\bibitem{Claessen_02} Claessen R, 
 Sing~M, Schwingenschl\"ogl~U, Blaha~P, Dressel~M
 and Jacobsen C~S
2002 \PRL \textbf{88} 096402

\bibitem{Klironomos_05} Klironomos A~D,
 Ramazashvili R~R and Matveev K~A
2005 \textit{Preprint} cond-mat/0504118

\end{thebibliography}
\end{document}